\documentclass[reprint,prb,superscriptaddress,showpacs]{revtex4-2}%
\usepackage{natbib} 

\setcitestyle{maxcitenames=6}
\usepackage{graphicx}
\usepackage{color}
\usepackage{amsmath}
\usepackage{amsfonts}
\usepackage{amssymb}
\usepackage{framed}
\usepackage{float}
\usepackage{placeins}
\usepackage[normalem]{ulem}
\usepackage{lineno}
\usepackage{lipsum}
\providecommand{\U}[1]{\protect\rule{.1in}{.1in}}
\makeatletter
\renewcommand*{\fnum@figure}{{\normalfont\bfseries \figurename~\thefigure}}
\renewcommand*{\@caption@fignum@sep}{\normalfont\textbf{ : }}
\makeatother
\usepackage{hyperref}
\hypersetup{
    colorlinks=true,
    linkcolor=blue,
    filecolor=magenta,      
    urlcolor=cyan,
}
\begin{document}

\title{ Altermagnetism in the layered intercalated transition metal dichalcogenide CoNb$_4$Se$_8$}

\author{Resham Babu Regmi}
\affiliation{Department of Physics and Astronomy, University of Notre Dame, Notre Dame, IN 46556, USA}
\affiliation{Stravropoulos Center for Complex Quantum Matter, University of Notre Dame, Notre Dame, IN 46556, USA}

\author{Hari Bhandari}
\affiliation{Department of Physics and Astronomy, University of Notre Dame, Notre Dame, IN 46556, USA}
\affiliation{Stravropoulos Center for Complex Quantum Matter, University of Notre Dame, Notre Dame, IN 46556, USA}
\affiliation{Department of Physics and Astronomy, George Mason University, Fairfax, VA 22030, USA}

\author{Bishal Thapa}
\affiliation{Department of Physics and Astronomy, George Mason University, Fairfax, VA 22030, USA}
\affiliation{Quantum Science and Engineering Center, George Mason University, Fairfax, VA 22030, USA}

\author{Yiqing Hao}
\affiliation{Neutron Scattering Division, Oak Ridge National Laboratory, Oak Ridge, Tennessee 37831, USA}

\author{Nileema Sharma}
\affiliation{Department of Physics and Astronomy, University of Notre Dame, Notre Dame, IN 46556, USA}
\affiliation{Stravropoulos Center for Complex Quantum Matter, University of Notre Dame, Notre Dame, IN 46556, USA}

\author{James McKenzie}
\affiliation{Department of Physics and Astronomy, University of Notre Dame, Notre Dame, IN 46556, USA}
\affiliation{Stravropoulos Center for Complex Quantum Matter, University of Notre Dame, Notre Dame, IN 46556, USA}

\author{Xinglong Chen}
\affiliation{Materials Science Division, Argonne National Laboratory, Lemont, IL, 60439, USA}

\author{Abhijeet Nayak}
\affiliation{Department of Physics and Astronomy, University of Notre Dame, Notre Dame, IN 46556, USA}
\affiliation{Stravropoulos Center for Complex Quantum Matter, University of Notre Dame, Notre Dame, IN 46556, USA}

\author{Mohamed El Gazzah}
\affiliation{Department of Physics and Astronomy, University of Notre Dame, Notre Dame, IN 46556, USA}
\affiliation{Stravropoulos Center for Complex Quantum Matter, University of Notre Dame, Notre Dame, IN 46556, USA}

\author{Bence G\'abor M\'arkus}
\affiliation{Department of Physics and Astronomy, University of Notre Dame, Notre Dame, IN 46556, USA}
\affiliation{Stravropoulos Center for Complex Quantum Matter, University of Notre Dame, Notre Dame, IN 46556, USA}

\author{L\'aszl\'o Forr\'o}
\affiliation{Department of Physics and Astronomy, University of Notre Dame, Notre Dame, IN 46556, USA}
\affiliation{Stravropoulos Center for Complex Quantum Matter, University of Notre Dame, Notre Dame, IN 46556, USA}

\author{Xiaolong Liu}
\affiliation{Department of Physics and Astronomy, University of Notre Dame, Notre Dame, IN 46556, USA}
\affiliation{Stravropoulos Center for Complex Quantum Matter, University of Notre Dame, Notre Dame, IN 46556, USA}

\author{Huibo Cao}
\affiliation{Neutron Scattering Division, Oak Ridge National Laboratory, Oak Ridge, Tennessee 37831, USA}

\author{J.F. Mitchell}
\affiliation{Materials Science Division, Argonne National Laboratory, Lemont, IL, 60439, USA}

\author{I. I. Mazin}
\affiliation{Department of Physics and Astronomy, George Mason University, Fairfax, VA 22030, USA}
\affiliation{Quantum Science and Engineering Center, George Mason University, Fairfax, VA 22030, USA}

\author{Nirmal J. Ghimire}
\email{Corresponding author: nghimire@nd.edu}
\affiliation{Department of Physics and Astronomy, University of Notre Dame, Notre Dame, IN 46556, USA}
\affiliation{Stravropoulos Center for Complex Quantum Matter, University of Notre Dame, Notre Dame, IN 46556, USA}

\date{\today}
\begin{abstract}
 \textbf{Altermagnets (AMs) are a new class of magnetic materials that combine the beneficial spintronics properties of ferromagnets and antiferromagnets, garnering significant attention recently. Here, we have identified altermagnetism in a layered intercalated transition metal diselenide, CoNb$_4$Se$_8$, which crystallizes with an ordered sublattice of intercalated Co atoms between NbSe$_2$ layers. Single crystals are synthesized, and the structural characterizations are performed using single crystal diffraction and scanning tunneling microscopy. Magnetic measurements reveal easy-axis antiferromagnetism below 168 K. Density functional theory (DFT) calculations indicate that A-type antiferromagnetic ordering with easy-axis spin direction is the ground state, which is verified through single crystal neutron diffraction experiments. Electronic band structure calculations in this magnetic state display spin-split bands, confirming altermagnetism in this compound. The layered structure of CoNb$_4$Se$_8$ presents a promising platform for testing various predicted properties associated with altermagnetism.}     
 \end{abstract}
 \maketitle

\section{\textbf{Introduction}}
Reducing  power consumption and developing materials for smaller and faster devices is one of the principal targets in the field of spintronics. Antiferromagnetic materials (AFMs) hold significant potential in this area because they offer greater stability and faster dynamics compared to ferromagnetic (FM) materials\cite{jungwirth2016antiferromagnetic}. However, unlike FMs, it is challenging to obtain an external response from AFMs, particularly from the  collinear ones, due to their zero net magnetization. Recent theoretical advancements have shown that, under specific crystallographic conditions, even the collinear AFMs can exhibit non-zero anomalous Hall effect, magneto-optical Kerr effect, and spin-splitting of electron bands, similar to FMs\cite{PNAS,hayami2019momentum,yuan2020giant,vsmejkal2020crystal}. These materials have been termed “altermagnets” (AMs)\cite{vsmejkal2022beyond,vsmejkal2022emerging,kmazin2022altermagnetism}.

In a collinear antiferromagnet, there exists at least one symmetry operation that maps one spin sublattice onto the other. Typically, this symmetry operation is a lattice translation or spatial inversion, which preserves the electron energy, resulting in Kramers (spin) degenerate electron bands, unlike in ferromagnets. However, in certain (so-called altermagnetic) cases this operation is neither translation nor inversion, but a mirror, glide, rotation, or similar operation. In these cases, the electron states at a general $k$-point are not spin-degenerate. At the same time, the sign of the spin  splitting alternates in momentum space, giving rise to the name ``altermagnets". As exciting as the new realization is, the symmetry requirements are quite stringent, and hence there are just a handful of materials identified as altermagnets. Some have been predicted theoretically\cite{vsmejkal2022emerging,vsmejkal2020crystal,vsmejkal2022beyond,gao2023ai}, while very few, such as MnTe and CrSb, have been realized experimentally\cite {gonzalez2023spontaneous,krempasky2024altermagnetic, lee2024broken, osumi2024observation, reimers2024direct}. Altermagnets combine the beneficial properties of both ferromagnets and antiferromagnets. For example, in momentum space, they feature spin polarized bands similar to those of FMs, while in real space, they exhibit collinear AFM ordering, resulting in resonance frequencies in the terahertz range, compared to the gigahertz frequencies typical of FMs. To this end, one of the most anticipated applications of AMs is in  terahertz tunnel junctions, which could significantly enhance the read/write speeds of electronic devices\cite{vsmejkal2022emerging}. Additionally, new theoretical proposals suggest that proximity of AMs to topological and superconducting materials could give rise to several interesting phenomena\cite{zhang2024finite,li2024realizing,lu2024varphi,maeda2024theory,giil2024quasiclassical,hong2024unconventional,cheng2024orientation,niu2024electrically,ouassou2023dc,maeland2024many,chourasia2024thermodynamic,wei2024gapless,zhu2023topological,banerjee2024altermagnetic,das2023transport,li2024creation}. This emerging field not only requires the discovery of new materials but also the development of easily exfoliable materials to create devices that could instantiate these phenomena. 

In this article, we report altermagnetism in a layered intercalated transition metal dichalcogenide (ITMD) compound CoNb$_4$Se$_8$. Single crystal samples are grown, and a hexagonal crystal structure with an ordered interstitial Co sublattice is verified by single crystal X-ray diffraction.  Magnetic, transport, and thermal properties measurements imply an easy-axis antiferromagnetic ordering. This spin direction and the A-type magnetic pattern are predicted by density functional theory (DFT) calculations and subsequently confirmed by single crystal neutron diffraction experiments. The DFT calculations yield an anisotropy energy of 0.7 meV per Co atom and a magnetic moment of 1.3 $\mu_B$, with the spin-flop field estimated to be on the order of 100 T. The calculated bands display a clear spin splitting in the compensated collinear antiferromagnetic state, consistent with the altermagnetic symmetry of this compound.

\section{\textbf{Results and Discussion}}

\begin{figure*}[ht!]
\begin{center}
\includegraphics[width=.9\linewidth]{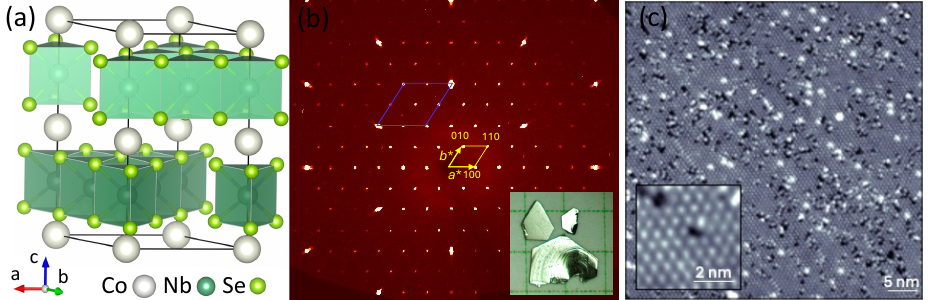}
    \caption{\small \textbf{Structure and structural characterization.} (a) Sketch of  crystal structure of CoNb$_4$Se$_8$. (b) Measured ($hk0$) zone of CoNb$_4$Se$_8$. The reciprocal unit cell of the ordered structure is outlined in yellow, while the  subcell corresponding to that of NbSe$_2$ base unit cell reported in Ref.\cite{VanLaar1971} is outlined in blue (also see Fig. S1). Inset shows an optical image of single crystals of CoNb$_4$Se$_8$. (c) STM topographic image of as-cleaved CoNb$_4$Se$_8$. Inset: a zoomed-in STM topographic image showing well-ordered Co triangular lattice.}
    \label{structure}
    \end{center}
\end{figure*}

It is well known that intercalating a 3d transition metal element into the 2H-Nb$X_2$ ($X$=S, Se) layers results in two distinct sets of ordered compounds\cite{VanLaar1971,Arita}. When a 3d transitional metal element occupies 1/4 of the octahedral holes, it forms an ordered centrosymmetric structure with a $2\times2$ superlattice based on the Nb$X_2$ unit cell in the hexagonal space group P6$_3$/mmc. In contrast, when the intercalant atom occupies 1/3 of the holes, it creates an ordered $\sqrt{3}\times\sqrt{3}$ superstructure within the non-centrosymmetric hexagonal space group P6$_3$22. CoNb$_4$Se$_8$ falls into the former category, and its schematic is depicted in Fig. \ref{structure}(a). We synthesized single crystals of CoNb$_4$Se$_8$ using a chemical vapor transport method (see Methods section for details) and determined the crystal structure through single crystal X-ray diffraction. The diffraction peaks obtained from a single crystal of CoNb$_4$Se$_8$ are shown in Fig. \ref{structure}(b), with an optical image of the crystals in the inset. The $2~\times~2$ superlattice unit cell, corresponding to the ordered structure in space group P6$_3$/mmc (associated with the 1/4$^{th}$ intercalation), is clearly visible in the diffraction peaks, highlighted by the yellow lines in Fig. \ref{structure}(b). The ordered structure contrasts with the original $1\times1$ disordered model\cite{voorhoeve1970intercalation}. Further details on the structural refinement can be found in Tables S1-S4. 

The structural characterization was further examined using STM topography, as shown in Fig. \ref{structure}(c), where a triangular lattice of Co atoms is observed, with clearer visualization in the inset. The distributed defects are likely due to the cleaving process. The $2\times2$ superstructure of the Co layer, with a lattice constant of 6.8 \AA, aligns with the measurements from single crystal X-ray diffraction and is confirmed by the Fourier transform presented in Fig. S2. In this structure, the bond center between two Co atoms along the $c$-axis lacks inversion symmetry due to the NbSe$_2$ prismatic layer, as illustrated in Fig. \ref{structure}(a). An A-type antiferromagnetic ordering (ferromagnetic Co planes coupled antiferromagnetically along $c$-axis, verified by our experiments and calculations presented below) in such a structure satisfies the criteria for altermagnetism\cite{PNAS,smolyanyuk2024tool,smolyanyuk2024tool2}. This was recently also pointed out for the isostructural and isomagnetic compound FeNb$_4$S$_8$\cite{lawrence2023fe}. While chemical substitution is necessary to stabilize FeNb$_4$S$_8$ in single crystal form, potentially affecting its band structure, unsubstituted CoNb$_4$Se$_8$ naturally forms single crystals with a well-ordered Co sublattice, offering a pristine platform for studying the altermagnetic properties.


\begin{figure*}[ht!]
\begin{center}
\includegraphics[width=.65\linewidth]{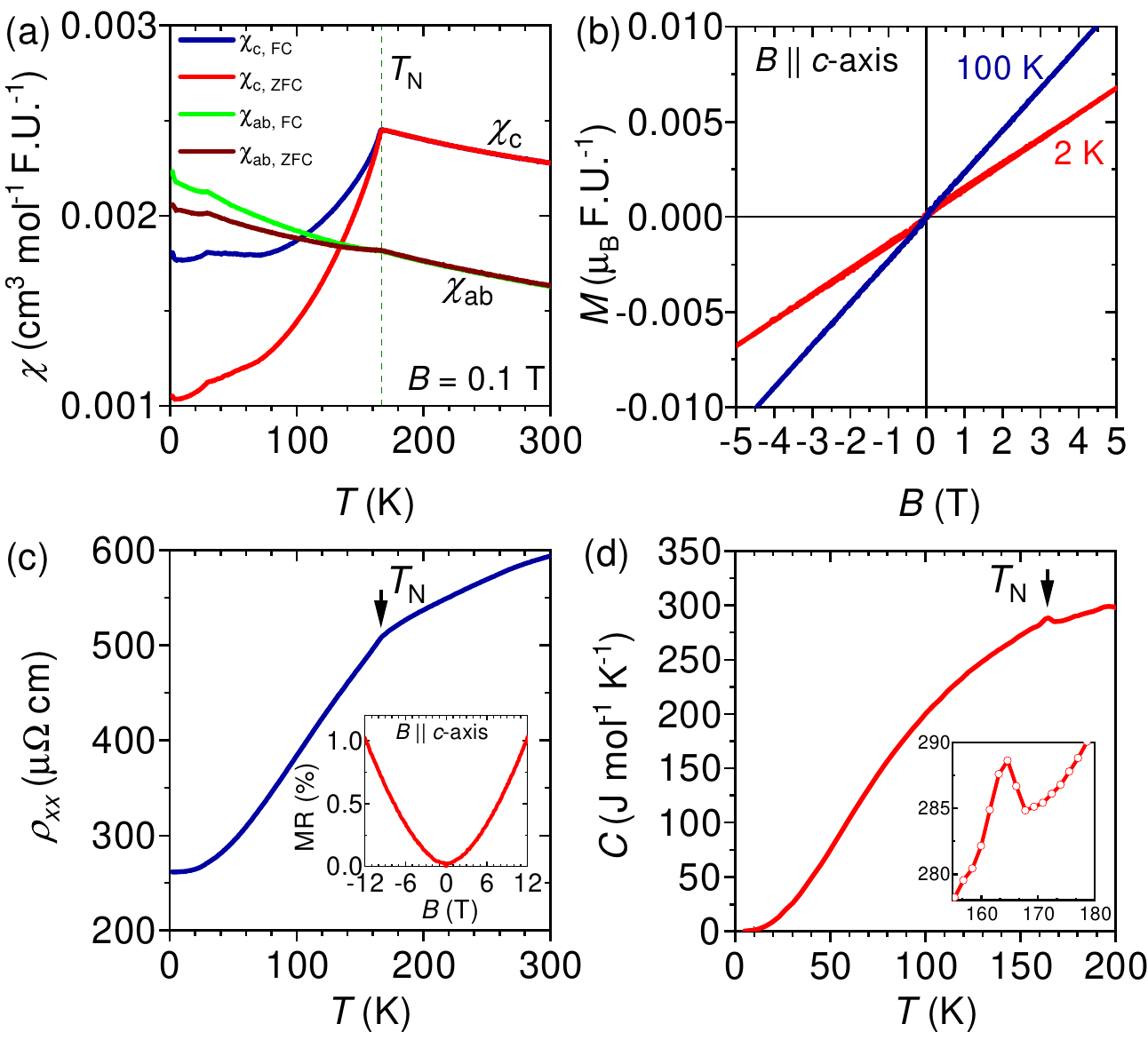}
    \caption{\small \textbf{Physical properties of CoNb$_4$Se$_8$.} (a) Magnetic susceptibility measured in a magnetic field of 0.1 T applied along $c$-axis ($\chi_c$) and within the $ab$-plane ($\chi_{ab}$). FC and ZFC represent the magnetic susceptibility measured with field-cooled and zero-field-cooled protocol, respectively. (b) Magnetization measured for the magnetic field applied along $c$-axis at two representative temperatures. (c) Electrical resistivity as a function of temperature. Inset shows the magnetoresistance at 1.8 K. (d) Heat Capacity as a function of temperature. Inset shows a magnified view of the lambda-type anomaly at T$_\text{N}$.}
    \label{suscept}
    \end{center}
\end{figure*}
Physical properties of CoNb$_4$Se$_8$ were measured using single crystals, as shown in Fig. \ref{suscept}. DC magnetic susceptibility ($\chi =M/B$), where $M$ represents the magnetic moment and $B$ is the external magnetic field, was measured with $B$ set at 0.1 T,  as depicted in Figure \ref{suscept}(a). When $B$ is aligned parallel to the $c$-axis, the susceptibility $\chi_{c}$ exhibits a sharp decrease below 168 K. In contrast, with $B$ in the $ab$-plane, there is a slight kink at 168 K in the susceptibility $\chi_{ab}$ followed by a gradual and modest increase at lower temperatures. This behavior is characteristic of antiferromagnetic ordering with moments aligned parallel to the $c$-axis\cite{blundell2001magnetism}. However, what is less typical is that $\chi$ remains large and anisotropic even above the N\'eel temperature ($T_{\text{N}}$).
We attribute this unusual behavior to the strong Stoner-enhanced Pauli susceptibility of the host material NbSe$_2$, as discussed in Supplementary section S2 and Fig. S3 

Furthermore, both $\chi_c$ and $\chi_{ab}$ exhibit a bifurcation between the field-cooled (FC), and zero-field-cooled (ZFC) data below $T_{\text{N}}$), likely due to domain walls freezing during the ZFC measurement. Additionally, a cusp-like feature in $\chi$ along  both directions is observed at 30 K. This feature may indicate either a second phase transition or the presence of a small impurity phase in the crystal, neither of which affects the conclusions presented here. The magnetization data $M(B)$, presented in Figure \ref{suscept}(b), is consistent with the antiferromagnetic ordering below 168 K. $M(B)$ at 2 and 100 K is linear with respect to $B$, with a small hysteresis observed between $\pm $2 T at 2 K (only as the broadening of the line), possibly linked to the 30-K susceptibility kink.

The magnetic transition at 168 K influences both the  electrical resistivity and heat capacity, as depicted in Figs. \ref{suscept}(c) and (d), respectively. Similar to its sulfide counterparts\cite{Parkin1980f,CoNb3S6, Ghimire2013,LIttle2020, nair2020electrical,Karna2021}, CoNb$_4$Se$_8$ is also a metal with a relatively high residual resistivity, a very small residual resistivity ratio (RRR) of 2.3, and a small magnetoresistance (MR) of $\approx$ 1 \% at 1.8 K and 12 T [see inset of Fig \ref{suscept}(a)].

\begin{figure*}[ht!]
\begin{center}
\includegraphics[width=1\linewidth]{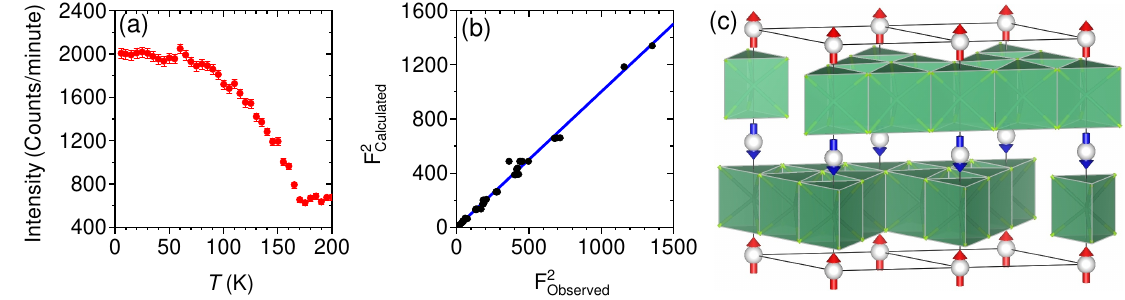}
    \caption{\small \textbf{Single crystal neutron diffraction and magnetic structure of CoNb$_4$Se$_8$.} (a) Magnetic ordering parameter measured at the magnetic Bragg peak at (1 0 1) showing the AFM transition temperature near T$_\text{N}$=168 K. (b) The observed squared structure factors $F^2_{\text{Observed}}$ versus calculated structure factors $F^2_{\text{Calculated}}$.The solid blue line is a guide to the eyes. (c) Magnetic structure determined by single crystal neutron diffraction at 5.6 K. Arrows represent the ordered magnetic moments.}  
    \label{neutron}
    \end{center}
\end{figure*}

After identifying the antiferromagnetic nature of the magnetic ordering, we conducted single crystal neutron diffraction experiments to investigate the microscopic characteristics of the magnetic ordering. Intensity of the magnetic Bragg peak at (1 0 1) as a function of temperature is shown in Fig. \ref{neutron}(a), which indicates a magnetic transition near $T_{\text{N}}$ = 168 K, consistent with the results from magnetic and transport measurements. Magnetic symmetry analysis using the Bilbao Crystallography Server\cite{perez2015symmetry} suggested two possible magnetic space groups (MSG):  P63/mm$^{\prime}$c$^{\prime}$ and P6$_{3^{\prime}}$/m$^{\prime}$m$^{\prime}$c. Among these, the MSG P6$_{3^{\prime}}$/m$^{\prime}$m$^{\prime}$c, corresponding to A-type magnetic order, provided a better fit to the observed data, as illustrated in Fig. \ref{neutron}(b). The lattice parameters obtained from neutron diffraction data are $a$ = $b$ = 6.904(14) \AA\ and $c$ = 12.321(11) \AA\ at 5 K, and $a$ = $b$ = 6.908(11) \AA\ and $c$ = 12.329(10) \AA\ at 200 K, which are in good agreement with those from X-ray diffraction ($a$ = 6.9178(13) \AA\ and $c$ = 12.390(3) \AA\ at 295 K).
The refined ordered moment of Co is 1.374(98) $\mu_B$, which aligns with the results from DFT calculations. The DFT calculations, including spin-orbit coupling (SOC), yield a spin moment of $\approx 1.35$ $\mu_B$ and an orbital moment of $\approx 0.11$ $\mu_B$, totaling $\approx 1.46$ $\mu_B$ per Co. It's worth noting that DFT calculations in good metals may slightly overestimate the ordered moment due to spin fluctuations. The itinerant nature of the magnetic moment on Co is further supported by the fact that other, less favorable magnetic patterns, such as two variants of the $q=(1/2,0,0)$ order or the ferromagnetic $q=(0,0,0)$ order, yield significantly reduced moments in the range of 0.9 - 1.2 $\mu_B$. 


\begin{figure*}[ht!]
\begin{center}
\includegraphics[width=1\linewidth]{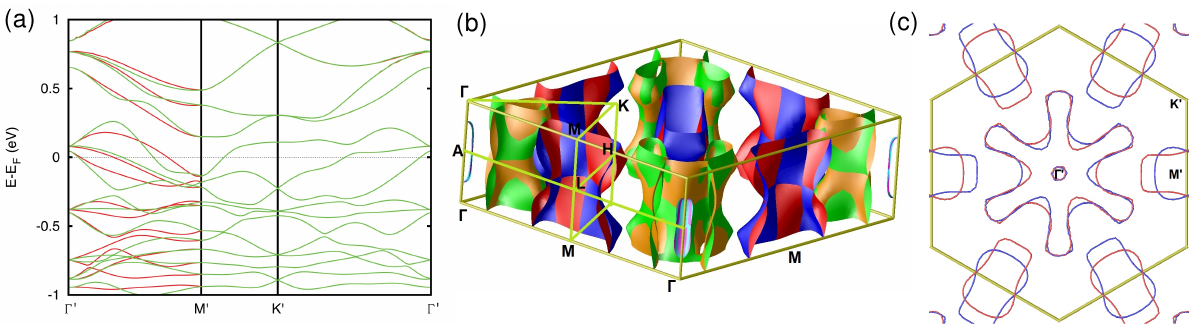}
    \caption{\small \textbf{Calculated electronic band structure and Fermi surface of CoNb$_{4}$Se$_{8}$.} (a) Calculated band structure of CoNb$_{4}$Se$_{8}$ with the A-type AFM (altermagnetic) structure. $\Gamma^{\prime}$, M$^{\prime}$ and K$^{\prime}$ points correspond to midpoints between $\Gamma$A, ML and KH, respectively. b) Fermi surface of the same. Green (brown), cyan (magenta)  and blue (red) colors correspond to the first (second) spin direction. Note the degeneracy planes, as described in the text. c) 
    The Fermi surface cut at $k_z=\pi/2c$, i.e., through the $\Gamma^{\prime}$-M$^{\prime}$-K$^{\prime}$ plane.}
    \label{bands}
    \end{center}
\end{figure*}

The A-type antiferromagnetic structure identified through neutron diffraction was also found to be the most energetically favorable according to DFT calculations (refer to Supplementary Section S5 for details of the calculations). As pointed out above, in this specific crystal structure, the A-type magnetic configuration exhibits altermagnetism. This is because the symmetry operation associated with spin reversal is a mirror, not an  inversion. The altermagnetic spin-split bands are clearly visible in the band structure calculations shown in Fig. \ref{bands}. Despite the presence of numerous symmetry-degenerate planes in the Brillouin zone (horizontal planes at $\Gamma$ and at A, and vertical planes passing through $\Gamma$-K and M-K lines), there is a significant induced spin splitting at intermediate $k_z$'s, which reveals the altermagnetism in CoNb$_4$Se$_8$. 

As shown in the Fermi surface plot in Fig. \ref{bands}, this material belongs to the class of $g$-wave altermagnet\cite{vsmejkal2022emerging}, with the aforementioned nodal planes parallel to $z$ and passing through $\Gamma$-K and K-M lines, as well as a horizontal nodal plane at $k_z$=0 and $\pi/c$. The Fermi surface consists of one hole pocket around $\Gamma$ and three compensating  electron pockets around M (neglecting a tiny hole pocket around $\Gamma$). As depicted in Fig. S4, at $T\sim$ 80 K, these pockets exhibit the same mobility and perfectly compensate each other in the ordinary Hall conductivity.

One manifestation of AM is the potential to exhibit an anomalous Hall effect (AHE). However, this effect is contingent on alignment of the spins (N\'eel vector) relative to the crystal structure. In CoNb$_4$Se$_8$, analyses\cite{smolyanyuk2024tool,smolyanyuk2024tool2} have shown that the only high-symmetry direction of the N\'eel vector that supports AHE is [210], meaning the easy axis must lie in the $ab$-plane and be perpendicular to the Co--Co bond. Since the magnetic easy axis in CoNb$_4$Se$_8$ is out-of-plane, the altermagnetic state does not permit AHE in any orientation, which aligns with the Hall resistivity data shown in Fig. S4.

However, the more significant aspect of AM, which is spin polarized bands, combined with the layered crystal structure of CoNb$_4$Se$_8$, could lead to many fascinating possibilities. The well-ordered Co sublattice observed in structural characterizations makes this compound promising. Based on our previous experience with two other ITMDs CrNb$_3$S$_6$\cite{Sirica2016}, and CoNb$_3$S$_6$\cite{Yang2021}, spin-split bands should be observable in photoemission experiments. Moreover, the ease of exfoliation into nanosheets makes it practical to interface the AM with other materials such as topological insulators and superconductors to test the proposed novel phenomena enabled by AM. The potential for epitaxial growth of the Se-based ITMDs\cite{litwin2023growth} also makes this system attractive for the fabrication and testing of anticipated AM tunnel junctions. Another possibility is to create heterostructures with other A-type or some other antiferromagnets with ferromagnetic planes, which could positively influence the N\'eel vectors and enhance the coupling of altermagnetic domains with an external field, as suggested in some studies\cite{mazin2023altermagnetism}. 

Additionally, it's important to note that Co intercalation leads to a noticeable lattice distortion in the NbSe$_2$ layer. As  illustrated in Fig. S6(a), Nb atoms farther away from Co are displaced, forming triangles with a side length of 3.295 \AA, while the other two Nb-Nb bonds have lengths of 3.460 and 3.623 \AA, respectively, with the longest bond about 0.33 \AA\ longer than the shortest. Simultaneously, the Se plane undergoes a vertical warping, where, somewhat counterintuitively, Se atoms above the smaller Nb triangles are pulled down, even closer to Nb, and the others are pushed up [see Fig. S6(b)]. Similar results are consistently observed in DFT optimization of the crystal structure. These distortions can be described as a $3\times 3$ charge density wave in both the Nb and Se planes. An intriguing, yet unexplored question is the potential phonon-magnon coupling - specifically, how these atomic displacements might affect the exchange interactions and, conversely, how magnetic order might affect the phonons.

\section{\textbf{Conclusion}}

 We successfully synthesized single crystals of CoNb$_4$Se$_8$, where the Co atoms occupy the 1/4 octahedral holes between NbSe$_2$ layers, resulting in a structure belonging to the space group P6$_3$/mmc. The A-type antiferromagnetic structure, characterized by easy-axis spin orientation as determined by neutron diffraction experiments, aligns well with the predictions from DFT calculations. These calculations also reveal spin-split bands characteristics indicative of altermagnetism. To establish altermagnetism, two criteria must meet: the crystal symmetry should ensure that bonds between any two antiferromagnetic atoms lack a center of symmetry, and there should be antiferromagnetic ordering with a zero propagation vector. CoNb$_4$Se$_8$ meets both criteria, and spin-split bands are clearly observed in the band structure calculations, confirming altermagnetism in this compound. Experimentally observing this should be straightforward, especially using spin-polarized ARPES experiments. The layered structure of CoNb$_4$Se$_8$ could open up new possibilities for testing various properties, particularly at interfaces with  phenomena such as ferromagnetism, antiferromagnetism, band topology, and superconductivity.

\section{\textbf{Methods}}\label{S2}

Single crystals of CoNb$_4$Se$_8$ were grown by chemical vapor transport using iodine as the transport agent. First, a polycrystalline sample was prepared by heating stoichiometric amounts of cobalt powder (Alfa Aesar 99.998\%), niobium powder (Alfa Aesar 99.8\%), and selenium pieces (Alfa Aesar 99.9995\%) in an evacuated silica ampule at 950 $^{\circ}$C for 5 days. Subsequently, 2 g of the powder was loaded together with 0.4 g of iodine in a fused silica tube of 14 mm inner diameter. The tube was evacuated and sealed under vacuum. The ampule of 10 cm length was loaded in a horizontal tube furnace in which the temperature of the hot zone was kept at 950 $^{\circ}$C and that of the cold zone was $\approx$ 850 $^{\circ}$C for 7 days. Several CoNb$_4$Se$_8$ crystals formed with a distinct, well-faceted flat plate-like morphology. 

The crystals of CoNb$_4$Se$_8$ were examined by single crystal X-ray diffraction (SC-XRD). The SC-XRD data were collected at room temperature using a Bruker D8 diffractometer equipped with APEX2 area
detector and Mo K$\alpha$ radiation ($\lambda$ = 0.71073 \AA). Data integration, cell refinement and numerical absorption corrections were performed by the SAINT program and SADABS program in APEX3 software\cite{APEX3,SADABS}. The precession images were also synthesized in APEX3. The structure was solved by Olex2 using direct methods with the XS structure solution program and refined with full-matrix least-squares methods on $F^{2}$ by the XL refinement package\cite{rj2009complete,sheldrick2008short}.

The STM measurements were performed in a Unisoku 1500 microscope at a temperature of 4.2 K in ultrahigh vacuum with SPECS Nanonics electronics. The crystals were cryogenically cleaved at a temperature of 80 K before inserting into the STM head. Gwyddion was used to process STM images.

Compositional analysis was done using an energy dispersive X-ray spectroscopy (EDS) at the Notre Dame Integrated Imaging Facility at room temperature. A Bruker EDS attached to a Magellan 400 - field emission scanning electron microscope (FESEM) was used for the measurement.  
  
DC magnetic susceptibility, and magnetization were measured using a Quantum Design magnetic property measurement system (MPMS XL 7-T). Resistivity,  magnetoresistance, Hall resistivity, and Heat capacity  measurements were performed in a 14-T Quantum Design Dynacool Physical Property Measurement System (PPMS). Single crystals of CoNb$_4$Se$_8$ were trimmed to adequate dimensions for electrical transport measurements. Crystals were oriented with the  [0 0 1] direction parallel to the applied field for the $c$-axis. Magnetic field and electric currents both in magnetic and magnetotransport measurements were applied to a random direction in the ab-plane. Resistivity and Hall measurements were performed using the 4-probe method. Pt wires of 25 $\mu$m were used for electrical contacts with contact resistances less than 30 Ohms. Contacts were affixed with Epotek H20E silver epoxy. An electric current of 2 mA was used for the electrical transport measurements. Contact misalignment in the magnetoresistance measurement was corrected by symmetrizing the measured data in positive and negative magnetic fields. Magnetic and magnetotransport measurements were verified in more than one crystals grown in more than one growth batches for reproducibility.

Single crystal neutron diffraction was carried out on the HB-3A DEMAND\cite{cao2018demand} at the High Flux Isotope Reactor at Oak Ridge National Laboratory. A wavelength of 1.533 \AA\ from a bent Si-220 monochromator\cite{chakoumakos2011four} was used for mapping the reciprocal space at selected temperatures and the data collection at 5.6 K and 200 K. The Bilbao Crystallography Server\cite{perez2015symmetry} was used for the magnetic symmetry analysis and Fullprof software\cite{rodriguez1993recent} for the nuclear and magnetic structure refinement.

{For the density functional (DFT) calculations a projector augmented wave method as implemented in the Vienna ab initio simulation package (VASP)\cite{VASP}. It was used both for structure optimization and for the search of the magnetic pattern. For the latter, a doubled supercell consistent with a hypothetical in-plane propagation vector (1/2,0,0), similar to that in CoNb$_3$S$_6$, was used. The results were fitted to a two nearest neighbors Heisenberg Hamiltonian, with excellent fit quality. The intraplanar exchange appears to be ferromagnetic, $J_{||}\approx 5.8$ meV, and interpalnar antiferromagnetic, $J_\perp\approx 25.6$ meV, normalized to a unit magnetic moment. In all cases a generalized gradient approximation for the exchange and correlation functional\cite{perdew1996generalized} was utilized. No LDA+U or other corrections beyond DFT were applied. Up to $11\times11\times 6$ $k$-point mesh (64 irreducible points) was used to structural optimization, and $48 \times 48\times 25$ for the Fermi surface and transport analyses. All calculations except those for magnetic anisotropy were performed without the spin-orbit coupling.

The optimized structure was then used with an augmented plane wave Wien2k code \textsc{WIEN2k}\cite{wien} for the Fermi surface analysis.

 Schematic of both crystal and magnetic structures were constructed using the three dimensional visualization system VESTA\cite{momma2011vesta}. 

\section{acknowledgments}
N.J.G., R.B.R., and I.I.M. were supported by  Army Research Office under Cooperative Agreement Number W911NF- 22-2-0173. H.B. acknowledges support from the NSF CAREER award DMR-2343536. Work at Argonne National Laboratory (single crystal X-ray diffraction) was supported by the U.S. Department of Energy, Office of Science, Basic Energy Sciences, Materials Science and Engineering Division. N. S. acknowledges support from a Materials Science and Engineering Fellowship. H.C. acknowledges the support from U.S. Department of Energy, Office of Science, Office of Basic Energy Sciences, Early Career Research Program Award KC0402020, under Contract No. DE-AC05-00OR22725. This research used resources at the High Flux Isotope Reactor, the DOE Office of Science User Facility, operated by Oak Ridge National Laboratory. X. L. acknowledges support from a Ralph E. Powe Junior Faculty Enhancement Award from ORAU. The authors thank Xinyu Liu for the help with measuring magnetic susceptibility.

\section{\textbf{References}}

%

\pagebreak

\widetext
\begin{center}
\pagebreak
\hspace{0pt}
\vfill
\textbf{\large Supplementary Information }
\vfill
\hspace{0pt}
\end{center}

\setcounter{equation}{0}
\setcounter{figure}{0}
\setcounter{table}{0}
\setcounter{page}{1}
\makeatletter
\renewcommand\thesection{SI\arabic{section}}
\renewcommand{\theequation}{S\arabic{equation}}
\renewcommand{\thetable}{S\arabic{table}}
\renewcommand\thefigure{S\arabic{figure}}
\renewcommand{\theHtable}{S\thetable}
\renewcommand{\theHfigure}{S\thefigure}

\section*{S1. Structural characterization}

\begin{table}[H]
\caption{Crystal data and structure refinement for CoNb$_4$Se$_8$. Numbers in parentheses are estimated standard deviation.}\label{T1}
\centering
\begin{tabular}{@{\hspace{.7cm}}l@{\hspace{.7cm}}l}	
 \hline
Empirical Formula&CoNb$_4$Se$_8$\\ 
 Crystal system         &              		  Hexagonal	$\hspace{5.0cm}$ \\ 
  Temperature (K)  & 295  \\ 
 $\lambda$ (\AA) &   0.71073 \\
  Space group & {\it 	P6$_3$/mmc} \\
 Unit cell $a$, $b$, $c$ (\AA) & 6.9178(13),\makeatletter~6.9178(13),~12.390(3)\\
 {\it V}, (\AA$^3$) & 513.5(2) \\ 
  {\it Z} & 2 \\ 
  Crystal size (mm$^3$) & 	0.15 × 0.08 × 0.01 \\ 
 $\rho_{\mathrm{calc}}$, g cm$^{-3}$ & 6.870 \\ 
 $\mu$, mm$^{-1}$ & 	34.201 \\ 
 F(000)	&926.0\\
 $2\theta$ range of data collection, deg & 6.578 to 60.902 \\ 
 Index ranges	&-9 $ \le$ h $ \le$ 9, -9 $ \le$ k $ \le$ 9, -17 $ \le$ l $ \le$ 17\\
 Reflections collected & 10583 \\ 
 Independent reflections & 338 [$R_{int}$ = 0.0378, $R_{sigma}$ = 0.0121] \\ 
 Data/restraints/parameters & 338/0/19 \\ 
 $R_1$,w$R_2$ [$F_0>4\sigma(F_0)$] & 0.0278, 0.0751 \\ 
 Goodness-of-fit on F$^2$	&1.186\\
Final R indexes [I$\ge$2$\sigma$ (I)]	&R$_1$ = 0.0212, wR$_2$ = 0.0642\\
Final R indexes [all data]	&R$_1$ = 0.0247, wR$_2$ = 0.0667\\
 Largest diff. peak and hole, e (\AA$^{-3}$) & 1.84/-1.17 \\
\hline
\end{tabular}
\end{table}

\begin{table}[H]
\caption{Fractional Atomic Coordinates (×10$^4$) and Equivalent Isotropic Displacement Parameters (Å$^2$×10$^3$) for CoNb$_4$Se$_8$. U$_{eq}$ is defined as 1/3 of the trace of the orthogonalised U$_{IJ}$ tensor. Numbers in parentheses are estimated standard deviation.}\label{T2}
\centering
\begin{tabular}{l@{\hspace{1.1cm}}l@{\hspace{1.1cm}}l@{\hspace{1.1cm}}l@{\hspace{1.1cm}}l@{\hspace{1.1cm}}l}								
		\hline	

        Atom    & $x$      &	  $y$	          &	      $z$	         &	  U(eq)	        \\
		\hline
 Co1& 0000 &	 0000 &	5000 &	9.1(3)\\
Nb1	& 0000 &	0000 &	7500 &	5.5(2)\\
Nb2	& 4921.0(4) &	5079.0(4) &	7500 &	5.31(18)\\
Se1	& 8315.8(2) &	6631.5(5) &	6171.9(4)	& 6.18(19)\\
Se2 &	3333 &	6667	& 8901.3(5) &	6.7(2)
\\ 
\hline      					
\end{tabular}
\end{table}

\begin{table}[H]
\caption{Calculated from DFT Fractional Atomic Coordinates (×10$^4$).}\label{TC}
\centering
\begin{tabular}
{l@{\hspace{1.1cm}}l@{\hspace{1.1cm}}l@{\hspace{1.1cm}}l@{\hspace{1.1cm}}l@{\hspace{1.1cm}}l@{\hspace{1.1cm}}l}										\hline										

        Atom    & $x$      &	  $y$	          &	      $z$	                \\
		\hline
 Co(2a)& 0000 &	 0000 &	5000 \\
Nb(2b)	& 0000 &	0000 &	7500 \\
Nb(6h)	& 4908 &	5092 &	7500 \\
Se(12k)	& 8326 &	6651 &	6167\\
Se(4f) &	3333 &	6667	&8917 \\
	    \hline
\end{tabular}

\end{table}

\begin{table}[H]
\caption{Anisotropic Displacement Parameters (Å$^2$×10$^3$) for CoNb$_4$Se$_8$. The Anisotropic displacement factor exponent takes the form: -2$\pi^2$[h$^2$a$^{*2}$U$_{11}$+2hka$^*$b$^*$U$_{12}$+…]. Numbers in parentheses are estimated standard deviation.}\label{T2}
\centering
\begin{tabular}
{l@{\hspace{1.1cm}}l@{\hspace{1.1cm}}l@{\hspace{1.1cm}}l@{\hspace{1.1cm}}l@{\hspace{1.1cm}}l@{\hspace{1.1cm}}l}										\hline											
     Atom & U$_{11}$ & U$_{22}$ & U$_{33}$ & U$_{23}$ & U$_{13}$ & U$_{12}$\\
\hline
Co1	& 8.3(4)	& 8.3(4)	& 10.9(6) &	0	& 0	& 4.14(18)\\
Nb1	&4.1(3)	&4.1(3)	&8.1(4)	&0	&0	&2.07(13)\\
Nb2	&3.8(2)	&3.8(2)	&8.0(3)	&0	&0	&1.58(16)\\
Se1	&5.5(2)	&5.2(2)	&7.8(3)	&0.46(10)	&0.23(5)	&2.62(12)\\
Se2	&6.1(2)	&6.1(2)	&7.8(3) &0	&0	&3.05(12) \\    					
	    \hline
\end{tabular}

\end{table}

\begin{figure}[H]
\begin{center}
\includegraphics[width=.7\linewidth]{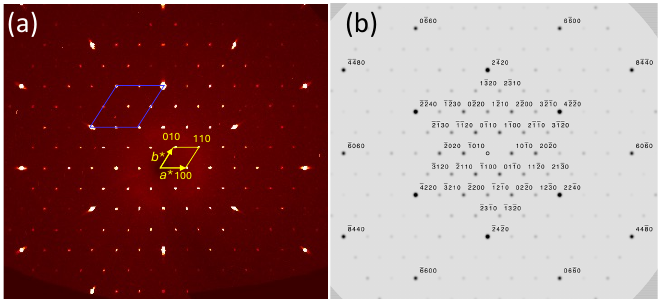}
    \caption{\small (a) Measured (\textit{hk}0) zone of CoNb$_4$Se$_8$. The reciprocal unit cell of the ordered structure is outlined in yellow, while the subcell of NbSe$_2$ is outlined in blue. b) Simulated (\textit{hk}0) zone based on the P6$_3$/mmc structure.}
    \label{FS}
    \end{center}
\end{figure}

\begin{figure}[H]
\begin{center}
\includegraphics[width=.8\linewidth]{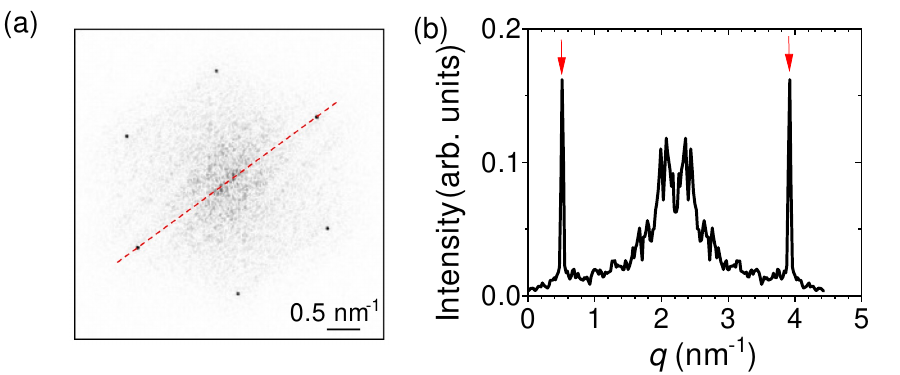}
    \caption{\small (a) Fourier transform of STM topographic image in Fig. 1c. (b) A line profile extracted along the dashed line in (a). Based on the strong Bragg peaks marked by red arrows, the lattice constant $a$ is 6.8 \AA.}
    \label{FS}
    \end{center}
\end{figure}

\section*{S2. Inverse susceptibility}
\begin{figure}[H]
\begin{center}
\includegraphics[width=.5\linewidth]{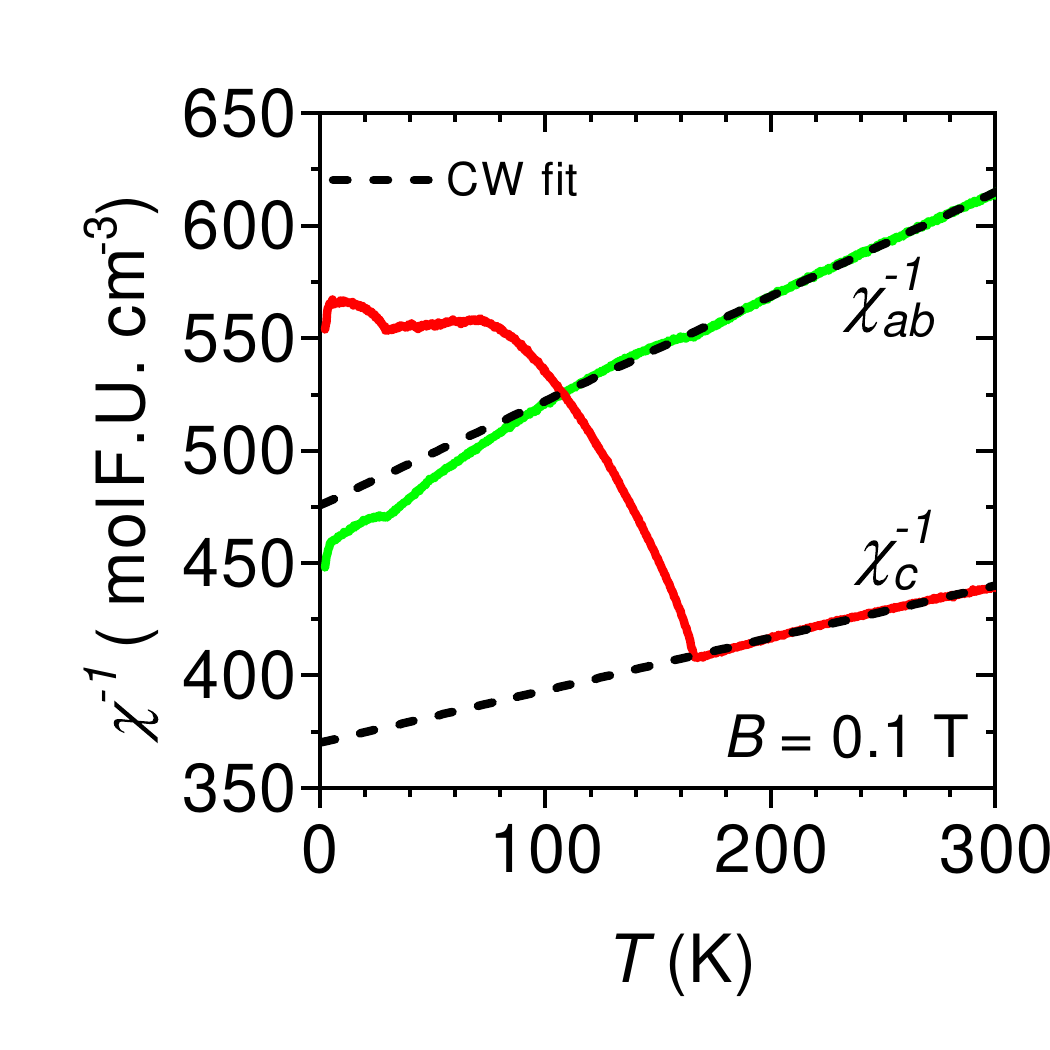}
    \caption{\small Inverse susceptibility of CoNb$_4$Se$_8$ measured with magnetic field of 0.1 T along $c$-axis ($\chi^{-1}_c$), and within $ab$-plane ($\chi^{-1}_{ab}$). The dashed lines are Curie-Weiss fit to the measured data.}
    \label{Invs}
    \end{center}
\end{figure}

Inverse susceptibility is shown in Fig. \ref{Invs}. An apparent Curie-Weiss (CW) behavior above $T_N$ is observed in both $\chi^{-1}_c$ and $\chi^{-1}_{ab}$. However, the intersection of $\chi^{-1}$ with the zero axis occurs at $-$300 -- $-$500 K, which is unphysically large (note that the net interplanar exchange coupling, $2J_\perp$ nearly cancels the intraplanar one, $-2J{\parallel}$), and the slopes give $\mu_{\text{eff}}$ of 4.1 -- 5.9 $\mu_B$/Co, also unphysically large. Note that our calculations and neutron diffraction experiment give, in agreement with chemical expectations, $M\sim 1.3$ $\mu_B$, corresponding to hybridization-reduced $S=1$, which should give  $\mu_{\text{eff}}\approx 2.8$ $\mu_B$, and negligible orbital moment for either magnetization direction. We interpret this odd result as a manifestation of the strong Stoner-enhanced Pauli susceptibility of NbSe$_2$. The latter was measured\cite{Karapetrov} to be $\approx 3\times 10^{-4}$ emu/mole, or  $\approx 1.2\times 10^{-3}$ emu/mole of (NbSe$_2)_4$, comparable to the total measured susceptibility at 300 K of about $2\times 10^{-3}$ (DFT calculations\cite{Ising} give an even larger $\chi\sim 4.2\times 10^{-4}$ emu/mole/NbSe$_2$). This Pauli susceptibility not only adds a temperature-independent background, but also renormalizes the Curie-Weiss susceptibility of Co ions, thus making any interpretation of the measured susceptibility in terms of the CW law meaningless.

\section*{S3. Hall Resistivity}

\begin{figure}[H]
\begin{center}
\includegraphics[width=.7\linewidth]{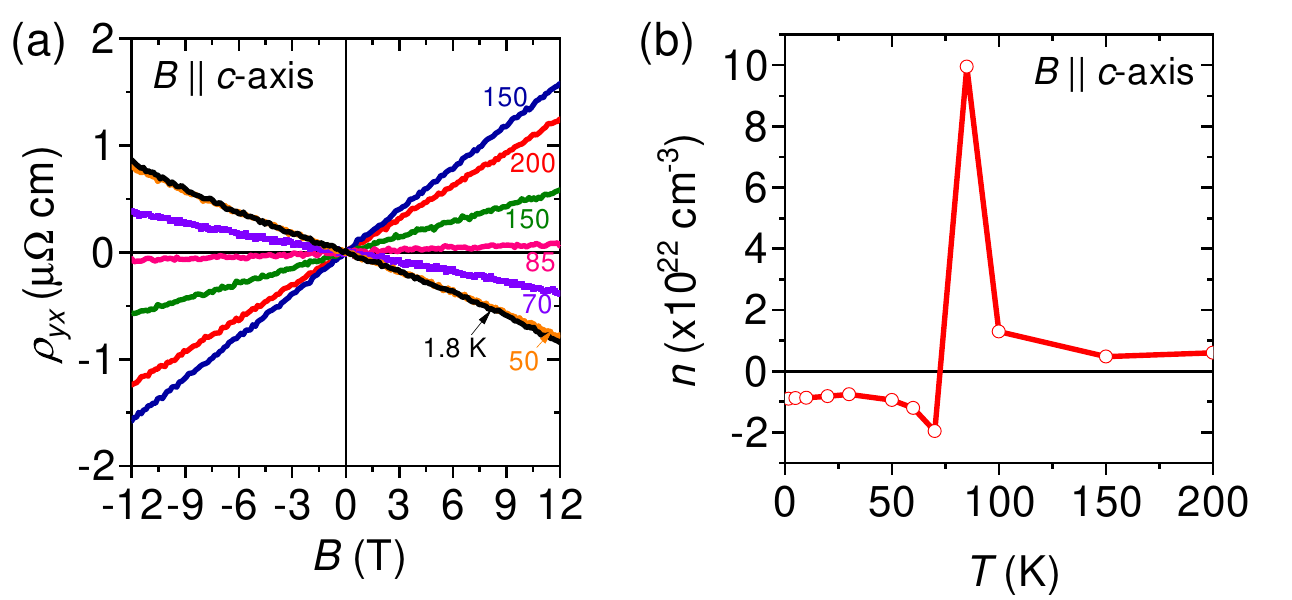}
    \caption{\small a) Hall resistivity of CoNb$_4$Se$_8$ measured at some selected temperatures. b) Formal carrier concentration of CoNb$_4$Se$_8$ as a function of temperature determined from Hall resistivity measurements within a single band model (since the Fermi surface, as discussed in the main text, consists of both hole and electron pockets, it is impossible to extract a physically meaningful number for the real carrier concentration). Note that at $T\sim 80$ K Hall hole and electron mobilities exactly compensate and the formal Hall carrier concentration diverges.}
    \label{FS}
    \end{center}
\end{figure}

\section*{S4. DFT calculations}

\begin{figure}[H]
\begin{center}
\includegraphics[width=.6\linewidth]{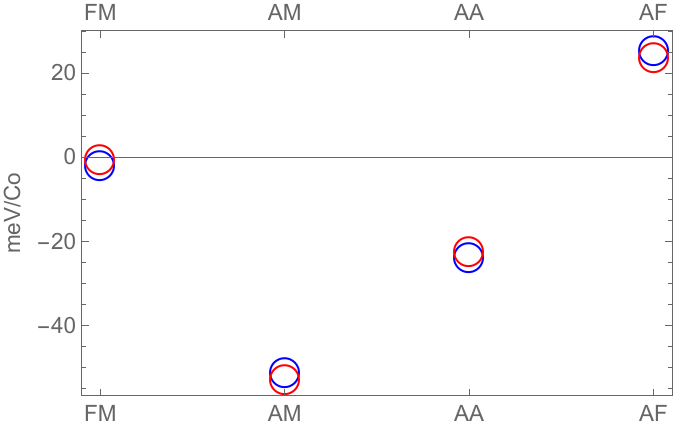}
    \caption{\small Calculated energies per Co with respect to the ferromagnetic configuration, for the four formula unit, compared to the two-exchange-constants Heisenberg fit. FM: ferromagnetic, AM: altermagnetic (ferro in plane, antiferro between the planes), AA: stripe order in the planes, antiferromagnetic between the planes, AF: stripe order in the planes, ferromagnetic between the planes. Note that the AM state is 30 meV/Co lower than any other. Red circles are
calculated data, and blue circles are the fitted ones.}
    \label{FS}
    \end{center}
\end{figure}

\pagebreak
\section*{S5. Structural distortion}

\begin{figure}[H]
\begin{center}
\includegraphics[width=0.8\linewidth]{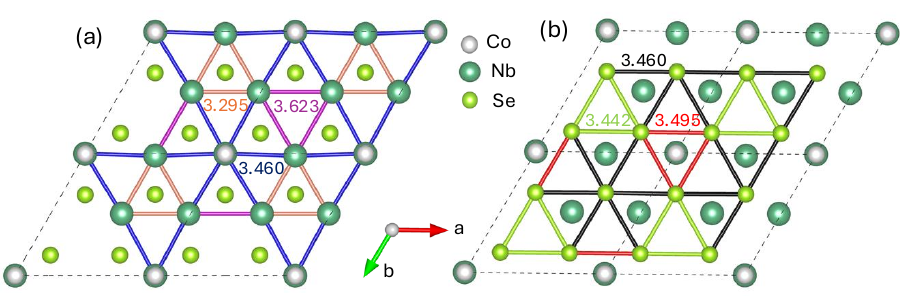}
    \caption{\small $C$-axis view of the crystal structure of CoNb$_4$Se$_8$, as determined by single crystal X-ray diffraction (described in Section S1), highlighting bonds in the Nb-layer (a), and Se-layer (b).  Bonds of different lengths are represented by cylinders in various colors. The bond lengths are labeled with their values in angstroms. The dashed lines indicate the crystallographic unit cells.}
    \label{FS}
    \end{center}
\end{figure}

\end{document}